\documentclass[a4paper,11pt]{article}
\usepackage{epsf,graphicx}
\usepackage{latexsym,amssymb,amsmath}
\usepackage{array}
\usepackage{setspace,cite}

\newcommand{\bddx}      {\ensuremath{\mathrm{Br(b\rightarrow D \overline{D}X)}}}
\newcommand{\bdd}      {\ensuremath{\mathrm{b\rightarrow D \overline{D}X}}}

\newcommand{\lnpj}     {\ensuremath{\mathrm{-ln}(P_j)}}
\newcommand{\nc}        {\ensuremath{n_c}}
 
\newcommand{\rphi}      {\ensuremath{\mathrm{{r}-{\phi}}}}

\newcommand{\bsl}       {\ensuremath{\mathrm{Br(b \rightarrow \ell \nu X)}}}
\newcommand{\bslups}   {\ensuremath{\mathrm{Br(B^{0/+} \rightarrow \ell \nu X)}}}
\newcommand{\zbb}       {\ensuremath{\mathrm{Z^0} \rightarrow b \overline{b} \;}} 
 
\newcommand{\nfrag}     {\ensuremath{\langle N_{\rm ch}\rangle_{\rm frag}}}

\begin{document}

\begin{titlepage}

\begin{center}{\large   EUROPEAN ORGANIZATION FOR NUCLEAR RESEARCH}
\end{center}\bigskip\bigskip
\begin{flushright}
       CERN-EP/2003-050   \\ 11 July 2003
\end{flushright}

\bigskip\bigskip\bigskip\bigskip\bigskip
\begin{center}
        {\huge \bf   A study of charm production in beauty decays with the OPAL detector at LEP}
\end{center}
\bigskip
\bigskip
\bigskip
\begin{center}
        {\LARGE \bf The OPAL Collaboration}
\end{center}
\bigskip
\bigskip
\bigskip
\bigskip
\bigskip
\begin{center}{\large  \bf Abstract}\end{center}

%
%%%%%%%%%%%%%%%%%      ABSTRACT     %%%%%%%%%%%%%%%%%%%%%%%%%%%%%%%%%%
%

The branching ratio of beauty hadrons to final states containing two charm hadrons,  \linebreak
$\bddx$, has 
been measured using an inclusive method in hadronic Z$^0$ decays 
with the OPAL detector at LEP.  The impact parameter significance of tracks 
opposite tagged b-jets is used to differentiate $\bdd$ decays from other decays.  The result is 
\begin{eqnarray*}
 \bddx = (10.0 \pm 3.2 ({\rm stat.})^{+2.4}_{-2.9}({\rm det.})^{+10.4}_{-9.0}({\rm phys.}))\%,
\end{eqnarray*}
where ``${\rm det.}$'' is the systematic uncertainty due to the modelling of the detector, and ``${\rm phys.}$'' is the systematic uncertainty due to the modelling of the underlying physics.  Using this result, the average number of charm plus anti-charm quarks produced in a beauty quark decay, $n_c$, is found to be $1.12^{+0.11}_{-0.10}$. 

\bigskip
\bigskip
\bigskip
\bigskip
\bigskip
\bigskip
\begin{center}{\large To be submitted to European Physical Journal C}    
\end{center}

\end{titlepage}

%%% Author List %%%%%%%%%%%%%%%%%%%%%%%%%%%%%%%%%%%%%%%%%%%%%%%%%%%%%%%%
\begin{center}{\Large        The OPAL Collaboration
}\end{center}\bigskip
\begin{center}{
%begin authorlist PLEASE DO NOT DELETE THIS COMMENT
G.\thinspace Abbiendi$^{  2}$,
C.\thinspace Ainsley$^{  5}$,
P.F.\thinspace {\AA}kesson$^{  3}$,
G.\thinspace Alexander$^{ 22}$,
J.\thinspace Allison$^{ 16}$,
P.\thinspace Amaral$^{  9}$, 
G.\thinspace Anagnostou$^{  1}$,
K.J.\thinspace Anderson$^{  9}$,
S.\thinspace Arcelli$^{  2}$,
S.\thinspace Asai$^{ 23}$
% ...
%end authorlist PLEASE DO NOT DELETE THIS COMMENT
}\end{center}\bigskip
\bigskip
%begin institutes
$^{  1}$School of Physics and Astronomy, University of Birmingham,
Birmingham B15 2TT, UK
\newline
$^{  2}$Dipartimento di Fisica dell' Universit\`a di Bologna and INFN,
I-40126 Bologna, Italy
\newline
$^{  3}$Physikalisches Institut, Universit\"at Bonn,
D-53115 Bonn, Germany
\newline
% ...
$^{  5}$Cavendish Laboratory, Cambridge CB3 0HE, UK
\newline
% ...
$^{  9}$Enrico Fermi Institute and Department of Physics,
University of Chicago, Chicago IL 60637, USA
\newline
% ...
$^{ 16}$Department of Physics, Schuster Laboratory, The University,
Manchester M13 9PL, UK
\newline
% ...
$^{ 22}$Department of Physics and Astronomy, Tel Aviv University,
Tel Aviv 69978, Israel
\newline
$^{ 23}$International Centre for Elementary Particle Physics and
Department of Physics, University of Tokyo, Tokyo 113-0033, and
Kobe University, Kobe 657-8501, Japan
\newline
% ...
$^{  *}$ Deceased

\newpage

%%%%%%%%%%%%%%%%%%%%%%%%%%%%%%%%%%%%%%%%%%%%%%%%%%%%%%%%%%%%%%%%%%%%%%%%%%%%
\section{Introduction} \label{sec:intro}
%%%%%%%%%%%%%%%%%%%%%%%%%%%%%%%%%%%%%%%%%%%%%%%%%%%%%%%%%%%%%%%%%%%%%%%%%%%%

Studying the decays of b hadrons\footnote{In this paper, Roman font ``b'' refers to the admixture of weakly decaying hadrons containing a beauty quark produced in e$^+$e$^-$ annihilations at $\sqrt{s} = {\rm m_{Z^0}}$; italic font ``$b$'' refers to the beauty quark. Weakly decaying hadrons containing a charm quark, $c$, are collectively called D hadrons.} allows important tests of the
Standard Model and Heavy Quark Effective Theory (HQET) to be made.  One such test is whether 
the average number of
$c$ plus $\overline{c}$ quarks produced in the decays of $b$ quarks, $\nc$, 
is consistent with theory.  The quantity $\nc$ can be determined 
experimentally by measuring the ``topological'' branching
ratios of b hadrons to different numbers of charm hadrons:
\begin{eqnarray}
  \label{eq:nc}
  \nc = 1 + \bddx + \rm{Br(b \rightarrow charmonium)} - \rm{Br(b
  \rightarrow no \; charm)}. 
\end{eqnarray}
This analysis measures the inclusive branching ratio of b hadrons to two 
charm hadrons, \linebreak 
$\bddx$, and combines this number with previous
measurements of Br(b $\rightarrow$ charmonium) and Br(b
$\rightarrow$ no charm) to obtain $\nc$. 
Neubert and Sachrajda use HQET to calculate $\nc = 1.20\pm0.06$\cite{neubert97}.  This theoretical prediction for $\nc$ is currently limited by uncertainty in the ratio of the charm and beauty quark masses ($0.25 < {\rm m}_c/{\rm m}_b < 0.33$).  

Besides being interesting in their own right, $\bddx$ and $\nc$ are 
correlated to the b hadron semi-leptonic branching ratio, 
$\bsl$. The current combined experimental values for $\bsl$ = (10.59 $\pm$ 0.22)\% at the Z$^0$ (LEP)\cite{pdg2002} and $\bslups$ = (10.38 $\pm$ 0.32)\% 
at the $\Upsilon$(4S) (CLEO)\cite{pdg2002} are slightly lower than the central values of theoretical predictions. QCD calculations within the parton model yield $\bsl > 12.5\%$\cite{bsl_theory,bigi94}, while more recent calculations that include radiative QCD corrections, spectator quark effects and charm quark mass effects yield $\bsl = 9.5\% \rightarrow 13.0\%$, depending on the renormalization scale and the quark mass scheme (pole or $\overline{\rm MS}$) used for the calculations \cite{bagan,neubert97}.  If any component of
the hadronic width ({\it e.g.} $\bdd$) is larger than expected, then the central value for $\bsl$ will be smaller.

The analysis presented in this paper makes the first inclusive
measurement of $\bddx$ and $n_c$ using OPAL data. This analysis uses a technique similar to one
employed by DELPHI \cite{delphi1}: a joint probability variable, constructed from track impact parameters, is used to
discriminate amongst the different b hadron decay topologies. 

%%%%%%%%%%%%%%%%%%%%%%%%%%%%%%%%%%%%%%%%%%%%%%%%%%%%%%%%%%%%%%%%%%%%%%%%%%%%
\section{The OPAL Detector and Data Samples} \label{sec:opal}
%%%%%%%%%%%%%%%%%%%%%%%%%%%%%%%%%%%%%%%%%%%%%%%%%%%%%%%%%%%%%%%%%%%%%%%%%%%%

A brief description of the most relevant components of the OPAL detector is given here; see Reference \cite{opalnim} for more details.  The central tracking system consisted of a silicon microvertex detector, a precision vertex drift chamber, a large volume drift (jet) chamber, and a set of chambers surrounding the jet chamber that made precise measurements of the $z$-coordinates of tracks\footnote{The OPAL coordinate system is right handed with the $z$-axis following the electron beam direction, and the $x$-axis pointing to the middle of the LEP ring. The azimuthal angle $\phi$ is measured in the $x-y$ plane with respect to the $+x$-axis and the polar angle $\theta$ is measured with respect to the $+z$-axis.}.
The silicon microvertex detector consisted of two layers of silicon strip detectors that provided three-dimensional hit information from 1993 onward.  The tracking system was located inside an axial 0.435 T magnetic field generated by a solenoidal coil just outside the tracking chambers.  Surrounding the solenoid were, in order, the scintillation time-of-flight detectors, the pre-sampling devices for the electromagnetic calorimeter, the lead glass electromagnetic calorimeter, the iron return yoke for the magnetic field (instrumented in order to provide hadron calorimetry), and farthest from the interaction point, the muon chambers.

The data used in this analysis were collected from 1993 to 1995. During these years, the centre-of-mass energy of the colliding beams, $\sqrt{s}$, was approximately $m_{{\rm Z^0}}$.  Data collected off the Z$^0$ peak are not used because the joint probability distributions used in this analysis vary as a function of centre-of-mass energy.  After applying standard OPAL hadronic Z$^0$ decay selection cuts \cite{z0_sel}, 1,866,000 events remain for analysis, with the contribution from background less than 0.1\%.

A total of 8 million simulated hadronic Z$^0$ decays are used to generate probability density functions (PDFs) for all signals and backgrounds considered for this analysis.   Of these decays, 3 million are $\zbb$ decays.  The simulated hadronic decays are generated by JETSET~7.4\cite{jetset}, using parameters tuned by OPAL \cite{opal_tune}, then processed using a full Monte Carlo (MC) simulation of the OPAL detector \cite{gopal}. Both real and simulated events are subjected to the same reconstruction and analysis algorithms.  

%%%%%%%%%%%%%%%%%%%%%%%%%%%%%%%%%%%%%%%%%%%%%%%%%%%%%%%%%%%%%%%%%%%%%%%%%%%%
\section{Method} \label{sec:method}
%%%%%%%%%%%%%%%%%%%%%%%%%%%%%%%%%%%%%%%%%%%%%%%%%%%%%%%%%%%%%%%%%%%%%%%%%%%%

\subsection{Joint probability} 
\label{sec:jntpbmethod}
\begin{figure}
  \begin{center}
\includegraphics[width=14cm]{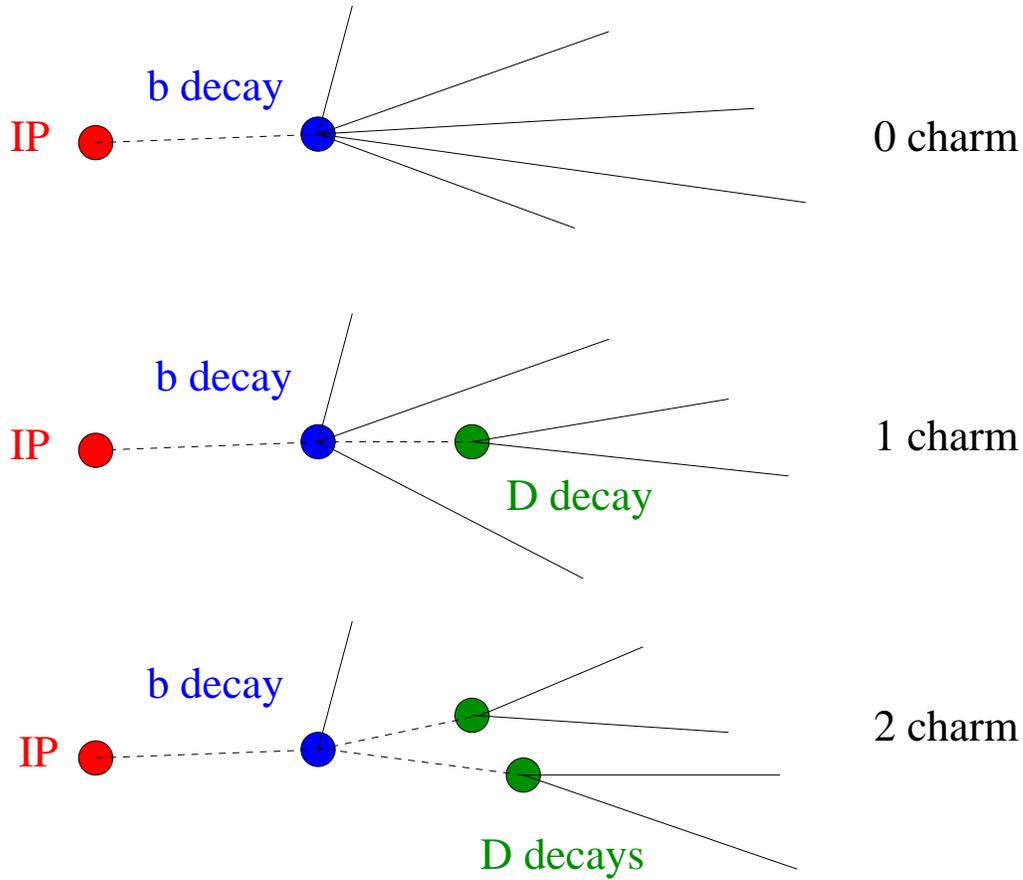}
  \end{center}
 \caption[]{Three different 
   topologies of b hadron decays.  As the charm hadron multiplicity increases, the tracks tend to 
originate farther from the interaction point (IP). Both b~$\rightarrow$~no~charm and 
b~$\rightarrow$~charmonium decays are represented by the upper 
   ``0 charm'' diagram. These two
   decay modes have similar topologies because of the prompt 
   electromagnetic or strong decays of charmonium states.}
 \label{fig:topologies}
\end{figure}

The main goal of this analysis is to differentiate double charm b hadron decays from 
single charm b hadron decays.  Single and double charm decays can be statistically separated by their different topologies. In b-jets, tracks from D hadron decays originate farther from the interaction point (IP) than tracks from b hadron decays.  Most of the tracks in $\bdd$ decays originate from D hadron decays, while most of the tracks in single charm b hadron decays originate from the location where the b hadron decays (see Figure \ref{fig:topologies}). 
The separation between a track and the IP (approximated by the reconstructed primary vertex) can be expressed in terms of its signed $\rphi$ impact parameter significance\footnote{The sign of $S$ is determined by the location in the $r-\phi$ plane where the track crosses the jet axis. The sign is positive (negative) if a line drawn from the primary vertex to the point where the track and jet axis cross points in the same (opposite) direction as the jet.}, $S = d_0/\sigma_{d_0}$, where $d_0$ is the impact parameter of the track with respect to the primary vertex in the $\rphi$ plane, and $\sigma_{d_0}$ is the uncertainty in this quantity. The $S$ values of the tracks in a jet are used to calculate a single variable for each jet: the joint probability, $P_j$. 

The joint probability is calculated by first considering the $S$ value of each selected track in a jet. Under the hypothesis that each track originates at the
IP, one can
calculate the conditional probability, $p_i$, for a track with $S>0$ to have the measured value $S$ or larger.  This probability is calculated
by integrating the measured $S$ resolution function for the OPAL detector, $f(S)$, beyond the measured value of $S$:
\begin{eqnarray}\label{eqn:pi}
  p_i = \frac {\int_{S_{\rm meas.}}^{S_{\rm cut}} f(S) dS} {\int_{0}^{S_{\rm cut}} f(S) dS}, 
\end{eqnarray}
where $S_{\rm cut}$ (= 25) is a cutoff in $S$ beyond which tracks are not considered.
Given an ensemble of tracks that originate from the interaction point, the distribution of $p_i$ will be uniform from 0 to 1.

The joint probability, $P_j$, is calculated by considering the $p_i$ 
of all the tracks in a jet:
\begin{eqnarray}
  P_j = y \sum_{m=0}^{N-1} \frac{(-\ln(y))^m}{m!}, 
\end{eqnarray} 
where $y$ is the product of the $N$ individual track probabilities.  
$P_j$ is the probability that the product of $N$ random numbers uniformly 
distributed
from 0 to 1 is $y$ or smaller.   The larger the $S$ of tracks in a jet, the 
smaller the $p_i$, and hence also $P_j$, will be. The possible values of $P_j$
go from 0 to 1, so $\lnpj$, the variable used in the analysis, varies from 0 
to $+\infty$.

$\bddx$ is measured by 
comparing simulated $\lnpj$ distributions of different b hadron decay 
topologies and backgrounds to data. The data distributions of $\lnpj$ are fit with the simulated $\lnpj$ distributions with $\bddx$ as a free parameter.

\subsection{Event, jet and track selection}  
\label{sec:selection}
%%%%%%%%%%%%%%%%%%%%%%%%%%%%%%%%%%%%%%%%%%%%%%%%%%%%%%%%%%%%%%%%%%%%%%%%%%%%

A number of event level cuts are applied to the data. The OPAL hadronic Z$^0$ decay selection is applied and the silicon microvertex detector is required to have been fully operational. Events are also required to be well contained in the
central portion of the tracking detectors (especially the silicon microvertex detector) by imposing a cut on the direction of the event thrust axis: $|\cos\theta_{T}| \leq 0.7$. In addition, events are required to be two-jet-like by applying a cut on the thrust of the event: $T>0.85$. The thrust and thrust axis of each event are calculated using tracks and energy clusters in the electromagnetic calorimeter that are not associated with any track.

All events are divided into two jets by the Durham jet finding algorithm \cite{durham}. The OPAL LEP-2 b-tagger\cite{opal_higgs_btag} is applied to these jets to select a sample of jets that is enriched in b hadron decays. This b-tagger has a good b-tagging efficiency while maintaining a high b hadron purity; this is achieved by combining track, high transverse momentum lepton, and jet shape information into a single likelihood variable.  For this analysis, the b-likelihood is required to be greater than 0.9 in order to
obtain a high purity b-jet sample (purity
$\sim$95\% and efficiency $\sim$40\%).   A jet is used in the analysis if the opposite jet passes the b-tag cut.  If both jets pass the b-tag cut then both jets are used. Using the opposite jet for b-tagging provides an unbiased sample of b-jets.  

Track selection cuts are made to select a sample of well measured tracks enriched in decay products of b and D hadrons.  As a preliminary track selection, tracks are required to have an $\rphi$ impact parameter with respect to the beam spot of less than 5 cm, at least 20 jet chamber hits, momentum $p<65.0$ GeV/$c$, and transverse momentum $p_T>0.15$~GeV/$c$. In addition, tracks are required to have $\rphi$ and $z$ coordinate hits in both layers of the silicon microvertex detector as MC studies show this requirement greatly reduces the systematic uncertainty due to detector resolution modelling. In order to reduce the number of fragmentation tracks, tracks are required to have a signed impact parameter significance, $S>0$, an angle with respect to the jet axis, $\theta_{t-j}<0.6$ radians, and a rapidity with respect to the jet axis, $y = \frac{1}{2}\ln(\frac{E+p_{||}}{E-p_{||}})>1.0$ ($E$ is the energy of the
track and $p_{||}$ is the component of the track's momentum
parallel to the jet axis). The efficacy of a $p$ cut for reducing the number of fragmentation tracks was investigated; however, MC studies show that a $p$ cut (beyond the $p_T>0.15$ GeV/$c$ requirement) resulted in an unnacceptably large increase in the statistical uncertainty of $\bddx$. The overall selection efficiency for tracks from b and D hadron decays is approximately 40\%.

\subsection{{\boldmath$S$} resolution function determination}
\label{sec:resfunc}

In order to calculate $P_j$, the $S$ resolution function of the OPAL detector, $f(S)$, must be known. The $S$ resolution functions are determined by fitting functions that are the sums of three gaussians plus an exponential, to  
tracks with $S<0$ (backward tracks) in a heavy flavour suppressed data sample (opposite jet has b-likelihood $<0.1$). Backward tracks are used as they tend not to have genuinely positive impact parameters; the $S$ distribution of backward tracks is dominated by detector resolution effects. Requiring b-likelihood $<0.1$ reduces the fraction of selected backwards tracks from b or D hadron decays from 11\% to 5\% (4\% from D decays). The presence of a small fraction of tracks from b and D hadron decays is not critical because all that is required is a sample of tracks to determine a common function, $f(S)$, that will be used to calculate $p_i$ and $P_j$ for data and MC.  The same event and track selection cuts are applied to select backward tracks for $f(S)$ and ``forward'' tracks for $P_j$ (except by definition, $S<0$ for backward tracks and $S>0$ for forward tracks).

\begin{figure}
  \begin{center}
\includegraphics[width=16cm]{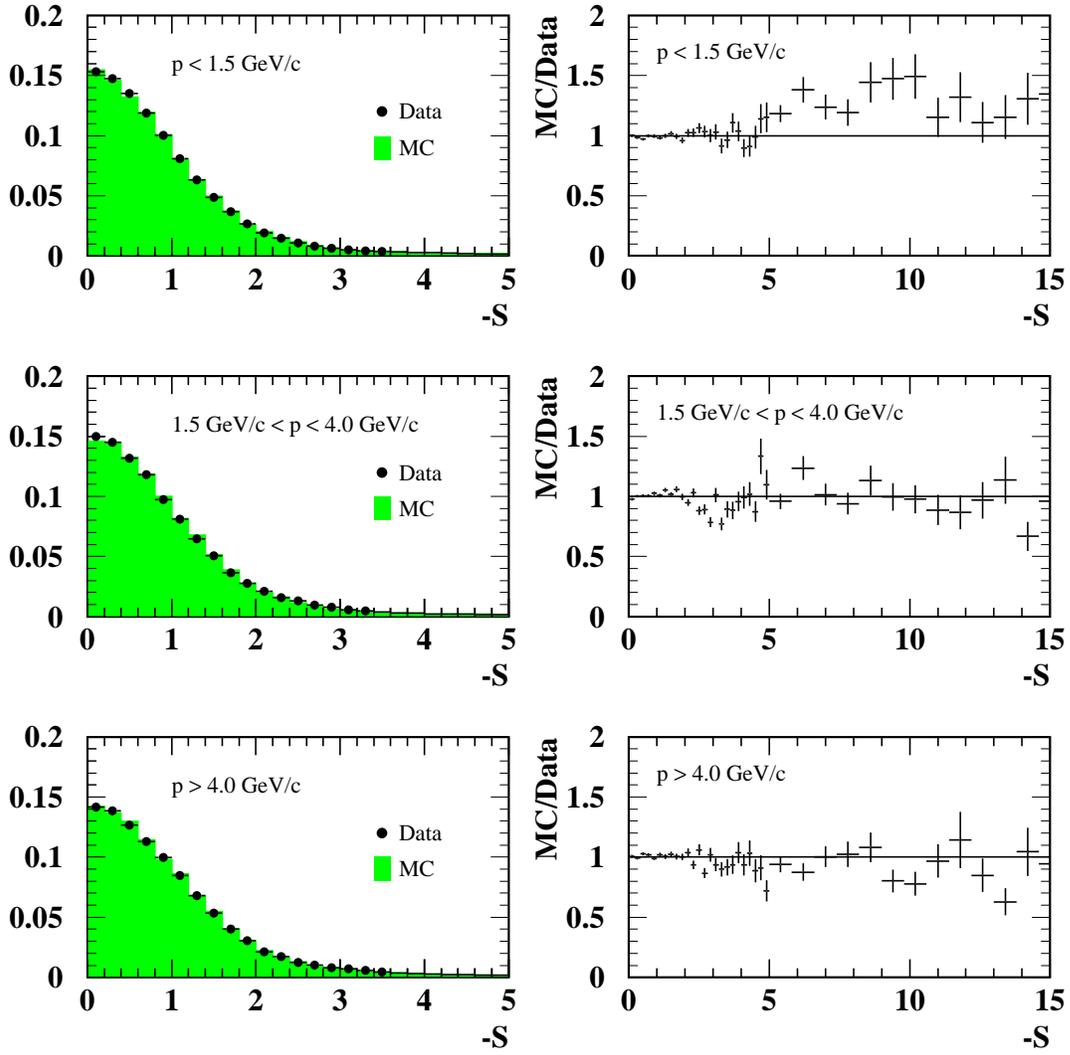}
  \end{center}
 \caption[Impact parameter significance distributions for
   SI-$r\phi$=SI-$z$=2 backward tracks]{Impact parameter significance distributions for backward tracks with hits in both layers of the silicon microvertex detector.  The plots in the left column show backwards tracks with $0<-S<5$. The data points with error bars represent the 1994 data and the histograms represent the corresponding simulation. The plots in the right column show the ratio of MC/data for $S$ values $0 < -S < 15$.}
 \label{fig:tuned_s}
\end{figure}

For each year of data taking, different $f(S)$ are determined for tracks in three momentum bins: $p<1.5 {\rm \;GeV}/c$, $1.5 {\rm \;GeV}/c < p < 4.0 {\rm \;GeV}/c$, and $p>4.0 {\rm \;GeV}/c$.  The agreement between simulated and real data for the $S$ distributions of backward tracks is shown in Figure \ref{fig:tuned_s}. The agreement is good for $S$ values less than 5 (where 99.7\% of the tracks lie); however, there are some differences in the high $S$ value tails of the distributions. The systematic uncertainty associated with this disagreement is discussed in Section \ref{sec:detmodel}.

%%%%%%%%%%%%%%%%%%%%%%%%%%%%%%%%%%%%%%%%%%%%%%%%%%%%%%%%%%%%%%%%%%%%%%%%%%

\subsection{Fitting procedure}
\label{sec:fitting}

A $\chi^2$ fit is performed to estimate the best fit
values of $\bddx$ for each year's data. The MINUIT package\cite{minuit} is used to minimize the $\chi^2$ function.
The fitting function used is
\begin{eqnarray}\label{eqn:fitting_func}
 F(x) =&N(1 +&\!\!\!\alpha x)\bigg[\{1-f_{uds}-f_{c}-f_{g}\}\times  \nonumber \\
%  &                       & \!\!\!\!\!\Big\{{\rm Br}_{0c}G^{0c}(x) + {\rm Br}_{1c}G^{1c}(x) + \nonumber  \\
%  &                       & \!\!\!     \;{\rm Br}_{2c}G^{2c}(x) + {\rm Br}_{\psi} G^{\psi}(x)\Big\} + \nonumber \\
  &                       & \!\!\!\!\!\Big\{{\rm Br}_{0c}G^{0c}(x) + {\rm Br}_{1c}G^{1c}(x) + {\rm Br}_{2c}G^{2c}(x) + {\rm Br}_{\psi} G^{\psi}(x)\Big\} + \nonumber \\
  &                       & \!\!\!\!\!\Big\{f_{uds}G^{uds}(x) + f_{c}G^{c}(x)+ f_{g}G^{g}(x)\Big\} \bigg]
\end{eqnarray}
where $x = -$ln$(P_j)$ and $N$ is a normalization factor chosen so that $\int \!F(x) dx$ is equal to the number of events in the data. The $G^i(x)$ are the normalized PDFs for the different signals and backgrounds. Br$_{1c}$ is the b $\rightarrow$ single charm branching ratio; Br$_{2c}$ is the double charm branching ratio; Br$_{0c}$ is the no charm branching ratio; Br$_{\psi}$ is the b$\rightarrow$ charmonium branching ratio. The background fractions are $f_{c}$, the fraction of charm jets, $f_{uds}$, the fraction of light quark jets, and $f_{g}$, the fraction of gluon jets from $\zbb$ events. Finally, $\alpha$ is a term used to parameterize mis-modelling due to incomplete knowledge of all physics inputs.   

MC studies show that changing the input values for various physics inputs such as the mean multiplicity of charged particles from fragmentation ($\nfrag$), b hadron lifetimes ($\tau_b$) or a variety of other inputs results in approximately linear changes in the $\lnpj$ distributions. 
The inclusion of the $\alpha$ term in Equation \ref{eqn:fitting_func} reduces the sensitivity of the analysis to changes in these physics inputs as this term allows the data to constrain these inputs. Consequently, several of the main systematic uncertainties are reduced by up to 50\% compared to using a fitting function without $\alpha$. This results in the total uncertainty being reduced by 30\%. 

Only Br$_{2c}$ and $\alpha$ are free parameters in the fit. Br$_{1c}$ is given by
\begin{equation}\label{constraint}
   {\rm Br}_{1c} = 1 - {\rm Br}_{2c} - {\rm Br}_{0c} - {\rm Br}_{\psi}, 
\end{equation}
while ${\rm Br}_{0c} =(0.7 \pm 2.1)\%$ \cite{delphi1}, and ${\rm Br}_{\psi} =(2.4 \pm 0.3)\%$ \cite{barker} are fixed. The background fractions $f_{uds}$, $f_{c}$ and $f_{g}$ are also fixed in the fit; the values of the fractions are determined from the MC.
 
The value of $\alpha$ is constrained to be close to zero by including a ``penalty function'' in the evaluation of the $\chi^2$ so that  
\begin{equation}
 \label{eqn:chi2}
 \chi^2 \rightarrow \chi^2 + \frac{\alpha^2}{\sigma_{\alpha}^2}.
\end{equation}
MC studies were performed to determine the value of $\sigma_{\alpha}$ in Equation \ref{eqn:chi2} that results in
the smallest predicted total uncertainty for $\bddx$: $\sigma_{\alpha}=3 \times 10^{-3}$.  This magnitude of $\sigma_{\alpha}$ corresponds approximately to the change in $\alpha$ when the sources of the main systematic uncertainties are varied by their one standard deviation uncertainties. The relative sizes of the statistical and systematic uncertainties changed appreciably as $\sigma_{\alpha}$ was varied in the MC studies; however, due to the anti-correlation between these uncertainties, the total uncertainty for $\bddx$ was not very sensitive to the value of $\sigma_{\alpha}$. Note that MC studies show that introducing the $\alpha$ parameter in the fit does not bias the estimator for $\bddx$ and reliable statistical uncertainties are estimated. 

\subsection{Binning of data by track multiplicity}

The data are divided into six track multiplicity (TM) bins to improve the sensitivity of the analysis and to reduce uncertainties due to incorrect modelling of the number of tracks contributing to $\lnpj$.  There is one bin for each value of track multiplicity in the range $1 \leq {\rm TM} \leq 5$ and one bin for ${\rm TM} \geq 6$. 

\begin{figure}
  \begin{center}
\includegraphics[width=16cm]{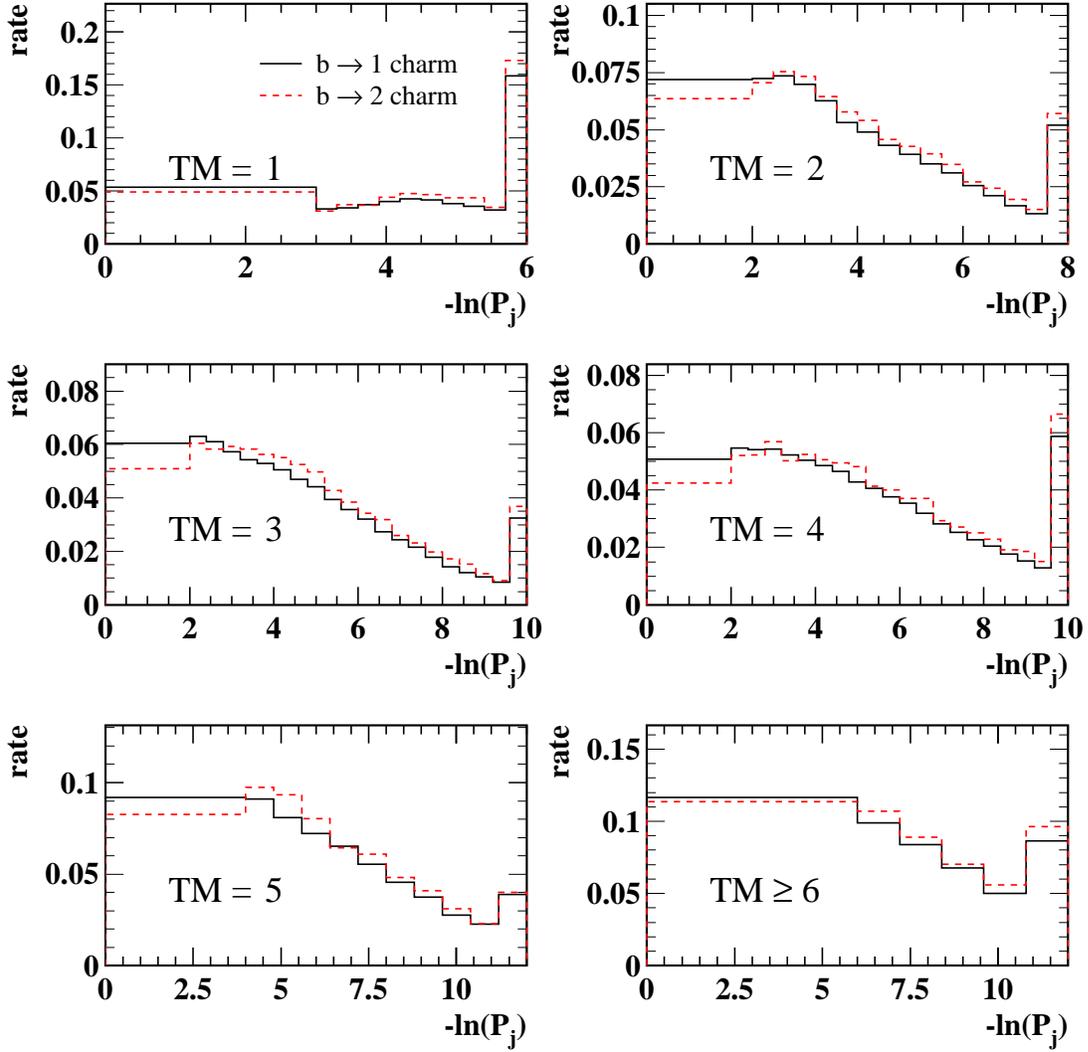}
  \end{center}
 \caption[]{The 1994 MC $\lnpj$ distributions for single and double charm b hadron decays for each track multiplicity bin. The last bin of each histogram includes all data beyond its upper edge. The solid and dashed histograms are for single and double charm b hadron decays, respectively.} 
 \label{fig:mc_pdfs}
\end{figure}

In order to ensure that the minimum number of expected data entries in each $\lnpj$ bin is at least 50, the number and range of $\lnpj$ bins varies for each track multiplicity.  Jets with $\lnpj$ greater than the range considered are included in the largest $\lnpj$ bin. The first five $\lnpj$ bins of each track multiplicity bin (ten bins in the case of TM=1) are also combined into one large bin in order to reduce the sensitivity of the analysis to changes in the $d_0$ resolution. Figure \ref{fig:mc_pdfs} shows the MC $\lnpj$ distributions for single and double charm b hadron decays for the different track multiplicity bins. 

Because the data are binned by track multiplicity, it is the fraction of $\bdd$ decays in b hadron decays for a particular track multiplicity bin, $f_{2c_i}$, that is actually determined in the fit for each bin.  According to the MC, $f_{2c_i}$ increases by a few percent (absolute) as the track multiplicity increases from one to six. $\bddx$ is determined by summing the results for all track multiplicity bins.   Using the fitted values of $f_{2c_i}$ for each bin and $N_{b_i}$, the total number of b hadron decays in each bin, the number of $\bdd$ decays ($= f_{2c_i}\times N_{b_i}$) is determined for each bin.  $\bddx$ is calculated for each year's data by dividing the total number of $\bdd$ decays by the total number of b hadron decays:  
\begin{equation}\label{total_b2c}
   \bddx = \frac{\sum_{i=1}^{6} f_{2c_i} N_{b_i}}{ \sum_{i=1}^{6} N_{b_i}}.
\end{equation}
Note that a single value of $\alpha$ is determined for all track multiplicity bins in a year.

%%%%%%%%%%%%%%%%%%%%%%%%%%%%%%%%%%%%%%%%%%%%%%%%%%%%%%%%%%%%%%%%%%%%%%%%%%%%
\section{Results} \label{sec:results}
%%%%%%%%%%%%%%%%%%%%%%%%%%%%%%%%%%%%%%%%%%%%%%%%%%%%%%%%%%%%%%%%%%%%%%%%%%%%

\subsection{Results for each year}

The results of the fits for each year are shown in Table \ref{tab:br2c}. The probabilities to obtain the $\chi^2$ values in Table \ref{tab:br2c} or larger for 84 degrees of freedom are 0.36, 0.16, and 0.59 for 1993, 1994 and 1995 respectively. 
The fitted values of $\bddx$ for each year and each track multiplicity bin are not constrained to the physically allowed region, $0\% < \bddx < 100\%$, because the results for each year and each track multiplicity bin are combined\cite{roos}.
\begin{table}[tbh!]
  \begin{center}
   \begin{tabular}{|c|r|c|r|c|}
    \hline
       year & data events & $\bddx$ (\%) & $\alpha$ ($\times 10^{-3}$) & $\chi^2$/d.o.f.  \\ \hline
       1993 &   408k & $-2.2 \pm 6.5$ & $1.0 \pm 2.4$ & 88.0/84 \\
       1994 & 1,076k & $15.0 \pm 4.4$ & $-0.8 \pm 2.0$ & 96.6/84 \\ 
       1995 &   382k & $15.5 \pm 6.7$ & $-1.4 \pm 2.7$ & 80.4/84 \\ 
    \hline
  \end{tabular}
  \caption[Results of $\lnpj$ fits for each year of data-taking]{Results of $\lnpj$ fits for each year of data-taking. The uncertainties are statistical only.}   
   \label{tab:br2c}
 \end{center}
\end{table}

The best fit values of $\alpha$ are consistent with zero for each year of data taking. This shows that the physics inputs to the MC and the detector modelling are in reasonable agreement with the data.  Due to differences in detector performance from year-to-year, it is not expected that the value of $\alpha$ should be the exactly same for each year; it is for this reason that $\alpha$ is determined separately for each year.
Note that the central value of $\bddx$ changes by less than one standard deviation when $\alpha$ is either unconstrained or omitted from the fitting function in fits to data or to a distinct sample of MC used instead of data.

\begin{figure}
  \begin{center}
\includegraphics[width=16cm]{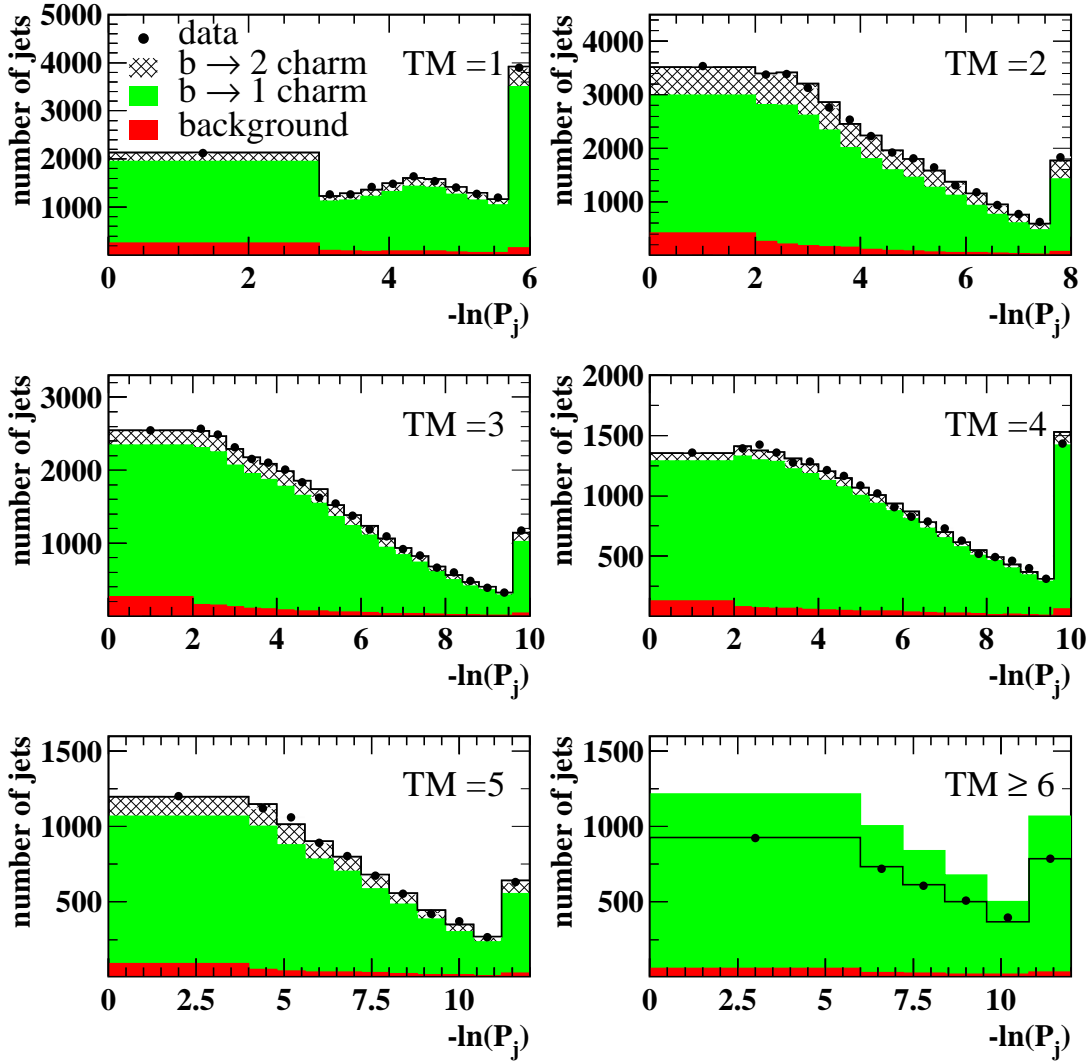}
  \end{center}
\vspace{-1cm}
 \caption[Fits of MC PDFs to data for each track multiplicity
bin]{Sum of fitted MC PDFs and data for all years for each track multiplicity
bin. The data points represent the data and the line histograms are the sums of the MC PDFs. 
The light grey histograms depict the single charm b hadron decays, and the hatched
histograms depict the $\bdd$ component.  The dark grey
histograms depict all backgrounds. The $\bdd$ contribution for each plot is calculated from the fitted 
$\bddx$ values for each track multiplicity bin from each year.}
 \label{fig:fit_combined}
\end{figure}

Figure \ref{fig:fit_combined} shows the sum of the fitted simulated PDFs and data for all years. 
For TM$\geq$6, the single charm component is fitted to be $>100\%$ and $\bddx < 0\%$; for the corresponding plot, 
there is no visible $\bdd$ component and the number of entries in the light grey histogram is greater 
than the number of entries in the line histogram (sum of all the MC PDFs).

A cross check of the analysis is performed by repeating the analysis for 1994 on a MC sample instead of data. The $\bddx$ result is $(17.4^{+5.5}_{-4.6}({\rm stat.}))$\%, which is consistent with the true $\bddx$ value in MC, 13.3\%. The fitted value of $\alpha$ is $(-0.02\pm 2.01)\times 10^{-3}$. 

\subsection{Combination of $\bddx$ results}

The $\bddx$ results for each year of data taking are combined to yield
\begin{equation}\label{comb}
  \bddx = (10.0 \pm 3.2 ({\rm stat.}) ^{+2.4}_{-2.9} ({\rm det.}))\%,
\end{equation}
where ``${\rm det.}$'' is the uncertainty due to detector modelling. The uncertainty due to detector modelling is considered to be uncorrelated from year-to-year as the detector modelling was tuned separately for each year. The size of this systematic uncertainty is approximately the same for each year.  Systematic uncertainties from the modelling of particle physics processes (see Section \ref{sec:physmodel}) are fully correlated from year-to-year so do not need to be considered when the weighted mean for $\bddx$ is calculated. The weights for each year are $1/(\sigma_{\rm stat.}^2 + \sigma_{\rm det.}^2)$.
The $\chi^2$/d.o.f. for combining the results from the three years is 3.7/2.  The probability to obtain this $\chi^2$ or larger with two degrees of freedom is 0.16. The $\bddx$ results can also be compared by track multiplicity bin. Note however, that it is expected that $f_{2c_i}$ should be slightly different for each bin.  The  $\chi^2$/d.o.f. for combining the results of each bin is 9.8/5. The probability to obtain this $\chi^2$ or larger with five degrees of freedom is 0.08.

%%%%%%%%%%%%%%%%%%%%%%%%%%%%%%%%%%%%%%%%%%%%%%%%%%%%%%%%%%%%%%%%%%%%%%%%%%%%
\section{Systematic Uncertainties} \label{sec:syst}
%%%%%%%%%%%%%%%%%%%%%%%%%%%%%%%%%%%%%%%%%%%%%%%%%%%%%%%%%%%%%%%%%%%%%%%%%%%%

The sources of systematic uncertainty investigated for this analysis are discussed in Sections \ref{sec:detmodel} (detector modelling) and \ref{sec:physmodel} (particle physics modelling).  Tables \ref{tab:systematics1} and \ref{tab:systematics2} summarize the systematic uncertainties.

\renewcommand{\arraystretch}{1.3}
\begin{table}[htbp]
  \begin{center}
    \begin{tabular}{|l|c|c|c|} \hline  
                       &   &  & Sign of \\         
   Source   & Value & $\sigma_{\bddx}$ (\%) & 
  $\frac{\Delta \bddx}{\Delta {\rm Source}}$ \\ \hline 
   $d_0$ modelling       & $\pm 1.5\%$  &  $^{+2.2}_{-2.8}$  & N/A  \\  %\hline
   $\sigma_{d_0}$ modelling &  $\pm 0.5\%$   & $\pm 0.5$ & N/A \\ %\hline
   $\epsilon_{\rm track}$       &  see text     &  $\pm 0.2$ & N/A   \\ %\hline
\hline
   Uncorrelated total &      & $^{+2.4}_{-2.9}$  &  \\ \hline
   $\nfrag$             &  $12.46 \pm 0.32$  &  $\pm 6.2$ & $+$ \\ %\hline 
   $\langle x_{{\rm b\rightarrow D}}\rangle$  & see text &  $\pm 0.5$ & N/A  \\ %\hline 
   $\tau_{\rm B^0}$         &  ($1.542 \pm 0.016$)ps &  $\pm 0.9$ & $-$ \\ %\hline
   $\tau_{\rm B^+}$         &  ($1.674 \pm 0.018$)ps &  $\pm 1.0$ & $-$  \\ %\hline
   $\tau_{\rm B_s}$         &  ($1.461 \pm 0.057$)ps &  $^{+0.7}_{-0.6}$ & $-$  \\ %\hline
   $\tau_{\Lambda_b}$       &  ($1.208 \pm 0.051$)ps &  $^{+1.0}_{-0.9}$ & $-$  \\ %\hline 
   $\tau_{\rm D^+}$         &  ($1.051 \pm 0.013$)ps &  $\pm 0.1$ & $-$  \\ %\hline 
   $\tau_{\rm D^0}$         &  ($0.412 \pm 0.003$)ps &  $\pm 0.3$ & $-$ \\ %\hline 
   $\tau_{\rm D_s^+}$       &  ($0.490 \pm 0.009$)ps &  $\pm 0.3$ & $-$ \\ %\hline 
   $\tau_{\Lambda_c^+}$     &  ($0.200 \pm 0.005$)ps &  $\pm 0.2$ & $-$ \\ %\hline 
   $f_{\Lambda_b}$          & ($10.5 \pm 2.0$)\%     &  $\pm 1.7$ & $+$ \\ %\hline 
   $f_{\rm B_s}$            & ($9.2 \pm 2.4$)\%      & $^{+0.4}_{-0.3}$ & $-$\\  %\hline
   $g\rightarrow b\overline{b}$ &  (2.54$\pm$0.50)$\times 10^{-3}$ & $\pm 0.0$ & N/A  \\  %\hline
   $g\rightarrow c\overline{c}$ &  (2.99$\pm$0.39)$\times 10^{-2}$ & $\pm 0.2$ & $+$   \\  \hline
   \end{tabular}
 \caption[Summary of systematic errors for $\bddx$ (part 1)]{Summary
    of systematic errors for $\bddx$ (part 1). The continuation of the
    summary, including the total correlated 
    uncertainty, is found in Table \ref{tab:systematics2}. The definitions and explanations of all the sources of uncertainty are contained in Sections \ref{sec:detmodel} and \ref{sec:physmodel}.}
\label{tab:systematics1}
 \end{center}
\end{table}
\renewcommand{\arraystretch}{1.0}

\renewcommand{\arraystretch}{1.3}
\begin{table}[htbp]
  \begin{center}
    \begin{tabular}{|l|c|c|c|} \hline  
                       &   &  & Sign of \\        
   Source                   & Value & $\sigma_{\bddx}$ (\%) &
    $\frac{\Delta \bddx}{\Delta {\rm Source}}$ \\ \hline 
   $\langle n_{\rm ch}\rangle_{{\rm D^+}}$   & $2.38 \pm 0.06$  & $^{+0.6}_{-0.8}$ & $-$ \\  %\hline
   $\langle n_{\rm ch}\rangle_{{\rm D^0}}$   & $2.56 \pm 0.05$  & $^{+0.9}_{-1.3}$  & $-$  \\  %\hline
   $\langle n_{\rm ch}\rangle_{{\rm D_s^+}}$ & $2.69 \pm 0.33$  & $^{+1.3}_{-1.0}$  & $-$ \\  %\hline
   $\langle n_{\rm ch}\rangle_{\Lambda_c^+}$ & $2.7 \pm 0.5$    & $^{+1.1}_{-0.9}$  & $-$ \\  %\hline
   $\langle n_{\pi^0}\rangle_{{\rm D^+}}$    & $1.18 \pm 0.33$  & $_{-0.9}^{+0.7}$    & $+$ \\  %\hline
   $\langle n_{\pi^0}\rangle_{{\rm D^0}}$    & $1.31 \pm 0.27$  & $^{+6.4}_{-3.5}$  & $+$   \\  %\hline
   $\langle n_{\pi^0}\rangle_{{\rm D_s^+}}$  & $2.0 \pm 1.4$    & $_{-1.4}^{+1.8}$  & $+$   \\  %\hline
   Br(D$^+ \rightarrow \overline{{\rm K}}^0$X) & $(61.2 \pm 7.8)\%$ & $^{+1.1}_{-1.0}$  & $+$ \\ %\hline
   Br(D$^0\rightarrow \overline{{\rm K}}^0$X) & $(45.5 \pm 5.9)\%$ & $\pm 1.3$  & $+$ \\  %\hline
   Br(D$_{\rm s}^+\rightarrow \overline{{\rm K}}^0$X)  & $(39^{+28}_{-27})\%$  &  $\pm 1.9$  & $-$ \\  %\hline
   Br($\Lambda_c^+ \rightarrow \overline{\Lambda}$X) & $(35 \pm 11)\%$ & $\pm 0.4$ & $+$ \\ %\hline 
   $\langle n_{\rm ch}\rangle_{{\rm b}}$     &  $4.97 \pm 0.07$  & $\pm 0.7$  & $-$ \\  %\hline
   $f_{{\rm D}^+}(1c)$      &  ($23.3 \pm 2.9$)\%    & $^{+1.3}_{-1.4}$   & $-$ \\  %\hline
   $f_{{\rm D}^+}(2c)$      &  ($17.0 \pm 4.9$)\%    & $_{-0.2}^{+0.3}$  & $-$ \\ %\hline 
   $f_{\Lambda_c^+}(1c)$    &  ($10.0 \pm 2.9$)\%    & $_{-2.4}^{+2.3}$  & $+$ \\ %\hline
   $f_{\Lambda_c^+}(2c)$    &  ($7.4 \pm 2.9$)\%     & $_{-1.3}^{+1.7}$   & $+$ \\ %\hline
   $\epsilon_g$             &  $\pm 10\%$     & $\pm 0.4$  & $+$ \\ %\hline
   $\epsilon_c$             &  $\pm 10\%$     & $\pm 2.0$  & $+$ \\ %\hline
   $\epsilon_{uds}$         &  $\pm 10\%$     & $\pm 0.8$  & $+$ \\ %\hline
   Br(b$\rightarrow$ no charm) & $(0.7 \pm 2.1)\%$ & $^{+3.0}_{-1.0}$ & $+$ \\ %\hline
   Br(b$\rightarrow$ charmonium) & $(2.4 \pm 0.3)\%$  &  $\pm 0.3$ & $+$ \\ \hline 
   Correlated total   &    & $^{+10.4}_{-9.0}$   &  \\  \hline 
   \end{tabular}
 \caption[Summary of systematic errors for $\bddx$ (part 2)]{Summary of systematic errors for $\bddx$ (part 2). The correlated systematic uncertainties are due to particle physics modelling.}
\label{tab:systematics2}
 \end{center}
\end{table}
\renewcommand{\arraystretch}{1.0}

\subsection{Detector modelling}\label{sec:detmodel}

Applying the same set of $f(S)$ to both data and MC to calculate $\lnpj$ 
assumes that the $S$ resolutions are the same between the two. The MC has been tuned to make the $d_0$ and $\sigma_{d_0}$ distributions of backwards
tracks in the MC match the data distributions (see Figure \ref{fig:tuned_s}). The MC $d_0$ distributions are tuned by scaling the difference between true and reconstructed track parameters.  The MC $\sigma_{d_0}$ distributions are tuned by scaling the values of $\sigma_{d_0}$ by a constant.

The uncertainty due to mis-modelling $d_0$ in the MC is determined by
re-processing the MC with the $d_0$ resolution varied by $\pm 1.5\%$.  This variation accounts for differences between the tuned MC track parameters determined by three different tuning methods. The MC is tuned by comparing MC and data backwards track $d_0$ distributions at small $d_0$ (width of core gaussian describing $d_0$ distribution) and over a wide range of $d_0$, and by comparing $\lnpj$ distributions for backwards tracks in a jet.

The uncertainty due to mis-modelling $\sigma_{d_0}$ is determined by repeating the analysis with $\sigma_{d_0}$ for each track scaled by $\pm$0.5\%. This variation accounts for remaining disagreement between the data and the MC.

After tuning the MC $d_0$ and $\sigma_{d_0}$ distributions, a difference remains between the tails of the data and MC $S$ distributions for backwards tracks (see Figure \ref{fig:tuned_s}). A study of the impact of this difference on the measured value of $\bddx$ was performed by varying, by 25\%, the fraction of tracks in the MC whose measured $d_0$ values are significantly different (greater than five standard deviations) from their true $d_0$ values. The measured value of $\bddx$ changes by a small amount (1.4\%) that is already covered by the systematic uncerainties attributed to $d_0$ and $\sigma_{d_0}$ modelling.

Different track selection efficiencies, $\epsilon_{\rm track}$, in the MC and the data may result in systematic differences between the MC and the data joint probability distributions. The most significant cut for $\epsilon_{\rm track}$ is the cut on the number of silicon microvertex detector hits associated with a track. After correcting the fraction of tracks in the MC with associated silicon hits (by randomly dropping tracks with silicon microvertex detector hits from the calculation of $\lnpj$), only a small statistical uncertainty remains.

\subsection{Particle physics modelling}\label{sec:physmodel}

The dominant sources of systematic uncertainty in this analysis are those which significantly affect the $S$ values of tracks included in the calculation of $\lnpj$.  The sources of the largest uncertainties for $\bddx$ are the charged particle multiplicity from fragmentation in $\zbb$ events, the neutral particle multiplicity of D decays, and the fractions of different D species in b hadron decays. Every correlated systematic uncertainty is calculated separately for each year, then combined for all years to yield the total systematic uncertainty due to each source. The total systematic uncertainty on $\bddx$ due to physics modelling is calculated by a quadrature sum of the individually combined systematic uncertainties.

Care must be taken when considering the systematic uncertainty due to 
fragmentation in $\zbb$ events, as $\langle x_{\rm E}\rangle$, the mean 
energy of weakly decaying b hadrons in Z$^0$ decays, is closely related to 
$\nfrag$, the average number of charged particles produced in the 
fragmentation process ({\it i.e.} charged particles not from the decay of the 
b hadrons).
Reweighting jets to vary $\langle x_{\rm E}\rangle$ changes $\nfrag$ at the same time. The contribution of 
each needs to be separated in order to avoid double counting systematic 
uncertainties. 

In order to determine the systematic uncertainty due to $\nfrag$ independently of $\langle x_{\rm E}\rangle$, $\nfrag$ is varied  by randomly dropping fragmentation tracks.  The mean multiplicity of charged particles from fragmentation in $\zbb$ events is determined by comparing experimental values of the average charged particle multiplicity in $\zbb$ decays, $\langle n_{\rm ch}\rangle_{b\overline{b}}$\cite{DeAngelis,opal_multbb}, and the average charged particle multiplicity of b hadron decays, $\langle n_{\rm ch}\rangle_{\rm b}$ (including the charged decay products of K$_{\rm S}$ and $\Lambda$)\cite{pdg2002}.  Combining these measurements gives $\nfrag = 12.46 \pm 0.32$.  The difference between $\bddx$ before and after dropping fragmentation tracks represents the systematic uncertainty.  The process of randomly dropping tracks is repeated 20 times to obtain a more precise estimate of the systematic uncertainty.

Several experiments have made precise measurements of 
$\langle x_{\rm E}\rangle$\cite{opal_bfrag,aleph_bfrag,sld_bfrag}. The combined result for the model-independent value of $\langle x_{\rm E}\rangle$ is $0.7151\pm0.0025$\cite{harder}. To assess the uncertainty in $\bddx$ due to uncertainty from $\langle x_{\rm E}\rangle$, 
jets originating from b hadrons in the simulation are reweighted so that the model-independent value of $\langle x_{\rm E}\rangle$ is varied by its one standard deviation uncertainty assuming the Bowler\cite{bowler}, 
Lund\cite{lundfrag}, and Kartvelishvili\cite{kart} fragmentation models.  

The change in $\bddx$ due to changing $\langle x_{\rm E}\rangle$ was found to be consistent with being due entirely to the associated change in $\nfrag$.   
This can be understood by realizing that the $d_0$ of tracks from b and D hadrons do not change significantly as a function of $\langle x_{\rm E}\rangle$; as $\langle x_{\rm E}\rangle$ increases, the decay lengths of the b and D hadrons increase but the opening angles of the tracks with respect to the jets decrease. These two effects tend to cancel out. For this reason, no systematic uncertainty is attributed to $\langle x_{\rm E}\rangle$ itself.  
  
CLEO has made precise measurements of the scaled\footnote{The momenta of the D hadrons are scaled by the maximum possible momentum for D hadrons produced in continuum e$^+$e$^-$ annihilations at $\sqrt{s}=10.58$~GeV.} momentum spectra of D$^{(*)}$ hadrons in B meson decays, $\langle x_{\rm b \rightarrow D}\rangle$ \cite{cleo_xbd}.  Those results are applied to the admixture of b hadrons produced in Z$^0$ decays. The uncertainty in $\bddx$ due to the uncertainty of $\langle x_{\rm b \rightarrow D}\rangle$ is determined by repeating the analysis many times with different, randomly generated MC $\langle x_{\rm b \rightarrow D}\rangle$ spectra that are compatible with the CLEO data. The width of the distribution of $\bddx$ values obtained with the different D momentum spectra determines the associated systematic uncertainty.

The lifetimes of the b and D hadrons are independently varied by their experimental uncertainties quoted in \cite{pdg2002}.  The fractions of different b hadrons are also varied as prescribed by the LEP Heavy Flavour Working Group \cite{lephfwg}.  

The $g \rightarrow b\overline{b}$ and $g \rightarrow c\overline{c}$ rates are varied by their experimental uncertainties \cite{lephfwg}. Requiring the thrust of events to be greater than $0.85$ greatly reduces this background, so the resulting systematic uncertainty for $\bddx$ is very small.

The Mark III collaboration has published values for the mean number of charged particles, $\langle n_{\rm ch}\rangle_{\rm D}$, and neutral pions, $\langle n_{\pi^0}\rangle_{\rm D}$, produced in D$^+$, D$^0$ and D$^+_{\rm s}$ decays \cite{mark3}. Br(D$\rightarrow \overline{{\rm K}}^0$X) was also measured.  These values, along with Br($\Lambda_c^+ \rightarrow \Lambda$X) \cite{pdg2002} and the charged particle multiplicity in charm baryon decays (varied $\pm 0.5$ about the JETSET prediction), are separately varied in the MC by their uncertainties to determine their contributions to the systematic uncertainty of $\bddx$. While each multiplicity is varied, the others are kept constant. Varying the neutral particle multiplicities affects the $p_T$ distribution of charged particles from D decays, which affects the $S$ distribution of tracks contributing to $\lnpj$.

The dependence of $\bddx$ on the mean charged particle multiplicity of b hadron decays, $\langle n_{\rm ch}\rangle_{\rm b}$, is reduced by binning the $\lnpj$ distributions by track multiplicity, but there still exists a systematic uncertainty due to $\langle n_{\rm ch}\rangle_{\rm b}$ \cite{lephfwg}.  The systematic uncertainty of $\bddx$ due to the uncertainty of $\langle n_{\rm ch}\rangle_{\rm b}$ is determined by reweighting jets to effectively change $\langle n_{\rm ch}\rangle_{\rm b}$ in the MC.  

The fractions of different D hadrons in single charm, $f_{D_i}(1c)$, and
double charm, $f_{D_i}(2c)$, b decays are varied because the different D hadron species have quite different lifetimes: the ratio of
$\tau_{\rm D^+} : \tau_{{\rm D}_{\rm s}^+} : \tau_{\rm D^0} : \tau_{\Lambda_c^+}$ is approximately 2.5 : 1.2 : 1 : 0.5. As the D$^+$ and
$\Lambda_c^+$ possess the longest and shortest D lifetimes, they are
the two D hadrons considered in this section.  Using measured rates of D hadron production in b hadron decays at the Z$^0$ \cite{pdg2002}, the following fractions are calculated: $f_{{\rm D}^+}(1c) = (23.3\pm2.9)\%$, $f_{{\rm D}^+}(2c) = (17.0\pm4.9)\%$, $f_{\Lambda_c^+}(1c) = (10.0\pm2.9)\%$, and $f_{\Lambda_c^+}(2c) = (7.4\pm2.9)\%$.

The backgrounds are divided into the following categories: gluon jets in Z$^0 \rightarrow b\overline{b}$ events ($f_g$), light quark background ($f_{uds}$), charm quark background ($f_c$), b $\rightarrow$ no charm decays, and ${\rm b \rightarrow charmonium}$ decays.  The fractions of backgrounds assumed to be present in the data are varied by their uncertainties to determine their contributions to the systematic uncertainty of $\bddx$. The $c$ and $uds$ selection efficiencies in b-tagging are varied by $\pm10\%$\cite{opal_btau}. The assumed branching ratios for ${\rm b \rightarrow charmonium}$ and b $\rightarrow$ no charm are also separately varied to assess their contributions to the systematic uncertainty.   

%%%%%%%%%%%%%%%%%%%%%%%%%%%%%%%%%%%%%%%%%%%%%%%%%%%%%%%%%%%%%%%%%%%%%%%%%%%%
\section{Conclusions} \label{sec:conclusion}
%%%%%%%%%%%%%%%%%%%%%%%%%%%%%%%%%%%%%%%%%%%%%%%%%%%%%%%%%%%%%%%%%%%%%%%%%%%%

The branching ratio $\bddx$ has been measured using an inclusive joint probability method with data collected by the OPAL detector at LEP.  The result 
\begin{eqnarray*}
 \bddx = (10.0 \pm 3.2 ({\rm stat.})^{+2.4}_{-2.9}({\rm det.})^{+10.4}_{-9.0}({\rm phys.}))\%  \end{eqnarray*}
is consistent with the inclusive measurement of $\bddx$ by DELPHI: $\bddx$ = ($13.6 \pm 3.0 ({\rm stat.}) \pm 3.0 ({\rm syst.})$)\%\cite{delphi1}.  Several significant sources of systematic uncertainty that are investigated in this analysis were not assigned uncertainties in the DELPHI analysis. These sources partially account for the difference in the size of the estimated systematic uncertainties.

Combining this $\bddx$ result with previous experimental
determinations of Br(b $\rightarrow$ charmonium)  $=(2.4 \pm 0.3)\%$ \cite{barker} and Br(b$ \rightarrow$ no charm) $=(0.7 \pm 2.1)\%$ \cite{delphi1} in Equation \ref{eq:nc} gives $n_c = 1.12^{+0.11}_{-0.10}$. This value is consistent with the average value of $n_c$ = $1.166 \pm 0.033$\cite{pdg2002}, from previous measurements carried out at the Z$^0$.  The measurements of $n_c$ and $\bddx$ obtained in this analysis are consistent with theoretical calculations\cite{neubert97}.
 
\par
\medskip
\noindent{\large \bf Acknowledgements}
\medskip
\par
\noindent We particularly wish to thank the SL Division for the efficient operation
of the LEP accelerator at all energies
 and for their close cooperation with
our experimental group.  In addition to the support staff at our own
institutions we are pleased to acknowledge the  \\
Department of Energy, USA, \\
National Science Foundation, USA, \\
Particle Physics and Astronomy Research Council, UK, \\
Natural Sciences and Engineering Research Council, Canada, \\
Israel Science Foundation, administered by the Israel
Academy of Science and Humanities, \\
Benoziyo Center for High Energy Physics,\\
Japanese Ministry of Education, Culture, Sports, Science and
Technology (MEXT) and a grant under the MEXT International
Science Research Program,\\
Japanese Society for the Promotion of Science (JSPS),\\
German Israeli Bi-national Science Foundation (GIF), \\
Bundesministerium f\"ur Bildung und Forschung, Germany, \\
National Research Council of Canada, \\
Hungarian Foundation for Scientific Research, OTKA T-038240, 
and T-042864,\\
The NWO/NATO Fund for Scientific Research, the Netherlands.\\

%%%%%%%%%%%%%%%%%%%%%%%%%%%%%%%%%%%%%%%%%%%%%%%%%%%%%%%%%%%%%%%%%%%%%%%
%   this is the Bibliography.  
%%%%%%%%%%%%%%%%%%%%%%%%%%%%%%%%%%%%%%%%%%%%%%%%%%%%%%%%%%%%%%%%%%%%%%

\end{document}